\documentclass[twocolumn,superscriptaddress,aps,prb,floatfix]{revtex4-2}

\usepackage{color}
\usepackage{graphicx}
\usepackage{amsmath,amssymb}
\usepackage{xcolor}
\usepackage{float}
\usepackage[colorlinks=true,linkcolor=blue,citecolor=blue,urlcolor=blue]{hyperref}

\graphicspath{{./}{figs/}}

\begin{document}
\pagestyle{plain}

\title{Ground-State Manifold Curvature and Phase Transitions in the Kitaev Honeycomb Model}

\author{Omid Yahyayi Monem}
\affiliation{Department of Physics, University of Mohaghegh Ardabili, P.O. Box 179, Ardabil, Iran}
\author{Setare Bashari}
\affiliation{Department of Physics, University of Mohaghegh Ardabili, P.O. Box 179, Ardabil, Iran}
\author{Hosein Mohammadzadeh}
\email{mohammadzadeh@uma.ac.ir}
\affiliation{Department of Physics, University of Mohaghegh Ardabili, P.O. Box 179, Ardabil, Iran}

\begin{abstract}
The Kitaev honeycomb model is an exactly solvable spin model whose ground state realizes a quantum spin liquid with gapped Abelian phases and a gapless phase that becomes a non-Abelian phase in a magnetic field. Here we study its phase transitions through the geometry of the ground-state manifold, using the Ricci scalar of the quantum metric. In the three gapped phases we obtain closed-form expressions for the curvature, which is negative and diverges at the critical lines; the three phases are related to one another by an exact change of coordinates. In the gapless phase, the curvature of the infrared-regularized metric is positive and decays inversely with the logarithm of the system size, with a size-independent amplitude that diverges at the phase boundaries, with an exponent different from that of the gapped side. The change of sign and the divergences together reproduce the complete phase diagram. A magnetic field gaps out the gapless phase and plays the role of the infrared cutoff, so that away from the critical lines the curvature of the non-Abelian phase becomes finite and positive, while around the critical lines it opens a crossover region whose width scales quadratically with the field. Since divergences of geometric quantities are often regarded as signatures of phase transitions, the divergence of the Ricci scalar at the critical lines provides a geometric signature of these transitions.
\end{abstract}

\maketitle

\section{Introduction}\label{sec:intro}

Quantum spin models are among the simplest settings in which strong
correlations, frustration, and quantum fluctuations combine to produce
rich collective behavior. From the Ising and Heisenberg chains to the XY
and XXZ models, they have served as laboratories for studying quantum
phase transitions, entanglement, and the breakdown of the Landau
symmetry-breaking paradigm
\cite{lieb1961two,pfeuty1970one,kargarian2008renormalization,jafari2008phase}. In recent
decades, attention has shifted toward models whose ground states cannot be
characterized by local order parameters, such as quantum spin liquids and
topologically ordered states \cite{wen1990topological,zhou2017quantum,savary2017quantum}.
In this context, the Kitaev honeycomb model occupies a special place
\cite{kitaev2006anyons}. It describes spin-$1/2$ degrees of freedom on a
honeycomb lattice with bond-dependent Ising interactions and is one of the
very few interacting two-dimensional quantum spin models that can be solved
exactly. Its ground state realizes a quantum spin liquid with fractionalized
excitations. The gapped phases are equivalent to the Abelian toric-code
phase, while in the gapless phase an applied magnetic field opens a gap and
produces non-Abelian Ising anyons, which are central to topological quantum
computation \cite{kitaev2003fault,nayak2008non}. The model has therefore been
studied extensively, and in connection with candidate Kitaev materials
\cite{jackeli2009mott,takagi2019concept}.

A complementary route to understanding quantum phase transitions is
provided by the geometry of the ground-state manifold. When the
Hamiltonian depends on a set of parameters, the ground state defines a
manifold whose geometry is encoded in the quantum geometric tensor
\cite{provost1980riemannian,berry1984quantal,kolodrubetz2013classifying}. Its imaginary part gives the Berry
curvature, while its real part is the quantum metric, which measures the
distance between neighboring ground states and is closely related to the
fidelity susceptibility and the quantum Fisher information
\cite{zanardi2006ground,zanardi2007information,campos2007quantum,zanardi2008quantum}.
Singularities and scaling behavior of the geometric phase, the fidelity, the
Loschmidt echo, and the quantum metric have been shown to signal quantum phase
transitions in a wide range of systems, including free-fermion models
\cite{carollo2005geometric,zhu2006scaling,carollo2020geometry,quan2006decay,cozzini2007quantum,einali2025quantum},
and the approach has been extended to thermal states through the Bures metric
\cite{zanardi2007bures}. Among these systems is the Kitaev honeycomb model
\cite{yang2008fidelity}, where the full quantum geometric tensor in the
presence of a magnetic field has recently been analyzed \cite{lu2025phase}.

The metric and the Berry curvature, however, are only the lowest-order
geometric objects of a parameter manifold. A fuller characterization comes
from its intrinsic curvature, encoded in the Riemann tensor and the Ricci
scalar derived from it. The Ricci scalar is used extensively in many branches of physics. It is
central to general relativity, and in thermodynamic and information geometry,
as well as in black-hole thermodynamics and quantum statistical mechanics, its
sign and divergences serve as indicators of interactions and critical behavior
\cite{weinhold1975metric,ruppeiner1979thermodynamics,janyszek1990riemannian,ruppeiner1995riemannian,quevedo2007geometrothermodynamics,mirza2009nonperturbative,mirza2010fractional,mohammadzadeh2026relativistic,ardabili2026alpha,monem2026intermediate,ardabili2026inclusion}.
For the Kitaev model, however, the Ricci scalar of the ground-state manifold
has not been studied systematically across the full phase diagram.

There are three reasons to prefer the scalar curvature over the components
of the metric as a probe of the phase structure. First, the metric components
depend on the choice of coordinates on the parameter manifold, so that their
form, and even the location of their singularities, can change under a
reparametrization; the scalar curvature is a coordinate invariant and
characterizes the intrinsic geometry of the ground-state manifold, which makes
it possible to distinguish genuine singularities from coordinate artifacts
\cite{kolodrubetz2013classifying}. Second, the metric is positive semidefinite
and carries no sign, whereas the scalar curvature can be positive or negative,
so that its sign can distinguish different phases. Third, it condenses the
three independent metric components of a two-parameter manifold into a single
scalar function, which can be mapped directly over the phase diagram. In the
Kitaev model, the components of the metric (equivalently, the fidelity
susceptibility) are already known to diverge at the phase boundaries
\cite{yang2008fidelity}; we show that the curvature provides additional,
coordinate-independent information, including a sign that differs between the
gapped and gapless phases.

In this paper, we use the Ricci scalar to investigate the phase transitions
of the Kitaev honeycomb model. Starting from the exact Majorana solution,
we reduce the ground state to a product of two-level states in momentum
space, so that the geometric tensor is determined entirely by the phase of
the Bloch function. We then use the Ricci scalar to study the phase
transition between the gapped and gapless phases. In the gapped phases the
Ricci scalar is obtained analytically, while in the gapless phase $B$ we
examine it for finite system size and analyze the finite-size effects.
Finally, we add the three-spin interaction generated by a magnetic
field, which gaps out phase $B$ into the non-Abelian phase, and show how the
field regularizes the curvature and modifies its behavior near the critical
lines.

The rest of the paper is organized as follows. Section~\ref{sec:geom}
reviews the geometry of the ground-state manifold. Section~\ref{sec:model}
introduces the Kitaev honeycomb model and its phase diagram, and
Sec.~\ref{sec:gs} presents its exact solution and ground state in the
vortex-free sector. In Sec.~\ref{sec:qgt}, we compute the quantum geometric
tensor and the quantum metric. Sections~\ref{sec:gapped} and
\ref{sec:gapless} give the metric and the Ricci scalar in the gapped phases
and in the gapless phase $B$, respectively. In Sec.~\ref{sec:map}, we use
the Ricci scalar to map the phase diagram and discuss its behavior near the
critical lines. Sec.~\ref{sec:field} and Sec. \ref{sec:scaling} analyze the effect of a magnetic
field and the scaling of the curvature near the critical line. In
Sec.~\ref{sec:discussion} we discuss the meaning and limitations of the
results. Finally, Sec.~\ref{sec:conc} summarizes our results.
Technical details are collected in the Appendix.

\section{Geometry of the Ground-State Manifold}\label{sec:geom}
In this section, we briefly review the geometry of the ground-state manifold, focusing on the quantum geometric tensor and its real and imaginary components. We then introduce the Ricci scalar as an intrinsic geometric quantity derived from the metric, which provides the basis for our subsequent analysis of quantum phase transitions.
Consider a manifold of Hamiltonians $H(\lambda)$ described by some coupling
parameters $\lambda=\{\lambda^\mu\}$. A natural measure of the distance between
the ground-state wave functions $|\psi_0\rangle$ separated by an infinitesimal
$d\lambda$ is~\cite{kolodrubetz2013classifying}
\begin{equation}\label{DS2}
ds^2 = 1-\left|\langle\psi_0(\lambda)|\psi_0(\lambda+d\lambda)\rangle\right|^2
= \sum_{\mu\nu}\chi_{\mu\nu}\,d\lambda^\mu d\lambda^\nu ,
\end{equation}
where $\chi_{\mu\nu}$ is the geometric tensor
\begin{equation}
\chi_{\mu\nu}=\langle\partial_\mu\psi_0|\partial_\nu\psi_0\rangle
-\langle\partial_\mu\psi_0|\psi_0\rangle\langle\psi_0|\partial_\nu\psi_0\rangle ,
\end{equation}
with $\partial_\mu\equiv\partial/\partial\lambda^\mu$. This tensor is invariant
under arbitrary $\lambda$-dependent $U(1)$ gauge transformations of the
ground-state wave function. Strictly speaking, Eq.~(\ref{DS2}) utilizes only the real
symmetric part of $\chi_{\mu\nu}$, which defines the metric tensor associated
with the ground-state manifold,
\begin{equation}
g_{\mu\nu}=\mathrm{Re}[\chi_{\mu\nu}]=\frac{\chi_{\mu\nu}+\chi_{\nu\mu}}{2},
\end{equation}
whereas its imaginary antisymmetric part is related to the Berry curvature through
\begin{equation}
\mathrm{Im}[\chi_{\mu\nu}]=-\tfrac12F_{\mu\nu}.
\end{equation}
In the present work, we focus on the real part of the quantum geometric tensor, which defines the quantum metric.

The Christoffel symbols, which act as connection coefficients, are expressed in
terms of the metric components and their derivatives as
\begin{equation}\label{Christoffel}
\Gamma^{i}_{jk}=\frac12\, g^{im}\left(\partial_k g_{mj}+\partial_j g_{mk}-\partial_m g_{jk}\right),
\end{equation}
where $g^{im}$ denotes the inverse metric components. The Riemann curvature
tensor, encoding the intrinsic curvature of the ground-state manifold, is
defined by
\begin{equation}\label{RiemannTensor}
R^{i}_{\;jkl}=\partial_k\Gamma^{i}_{lj}-\partial_l\Gamma^{i}_{kj}
+\Gamma^{i}_{km}\Gamma^{m}_{lj}-\Gamma^{i}_{lm}\Gamma^{m}_{kj}.
\end{equation}
From the Riemann tensor, the Ricci tensor is obtained via contraction,
\begin{equation}\label{RicciTensor}
R_{ij}=R^{m}_{\;imj},
\end{equation}
and the scalar curvature is given by
\begin{equation}\label{Ricci}
R=g^{ij}R_{ij}.
\end{equation}
The scalar curvature is a coordinate invariant, but it is not
invariant under a constant rescaling of the metric: under
$g_{\mu\nu}\to c\,g_{\mu\nu}$ one has $R\to R/c$. Its numerical values
therefore depend on the normalization of the metric, which we fix in
Sec.~\ref{sec:qgt}.
Singularities in the metric $g_{\mu\nu}$ and the scalar curvature $R$ identify points at which the ground state becomes non-analytic, signaling quantum phase transitions.

\section{Kitaev Honeycomb Model}\label{sec:model}
In this section, we briefly review the Kitaev honeycomb model and discuss the coupling conditions that determine its phase diagram. In particular, we summarize the relations between the coupling parameters at the phase boundaries separating the gapped and gapless phases. The Kitaev honeycomb model describes spin-$1/2$ degrees of freedom
residing on the vertices of a honeycomb lattice, with bond-directional
Ising-type interactions,
\begin{equation}\label{HKM}
H
=
-J_x\sum_{\langle ij\rangle_x}\sigma_i^x\sigma_j^x
-J_y\sum_{\langle ij\rangle_y}\sigma_i^y\sigma_j^y
-J_z\sum_{\langle ij\rangle_z}\sigma_i^z\sigma_j^z ,
\end{equation}
where $\sigma^\alpha_j$ are the Pauli matrices, and each site belongs to
exactly one link of each type $\alpha=x,y,z$ \cite{kitaev2006anyons}.

The importance of the model is threefold. First, it is one of the very few
interacting two-dimensional quantum spin models that is exactly solvable.
Second, its ground state realizes a quantum spin liquid, with fractionalized
excitations and no local order parameter. Third, in its gapped phases it is
equivalent to the toric code with Abelian anyons, whereas in the gapless phase,
upon applying a magnetic field, it acquires a gap and supports non-Abelian
Ising anyons, which are relevant for topological quantum computation.

The phase diagram is controlled by the relative magnitudes of the three
couplings. Taking $J_x,J_y,J_z\ge 0$ without loss of generality, the
model is in the gapless $B$ phase when the couplings satisfy all three
triangle inequalities,
\begin{equation}
J_x \le J_y + J_z, \qquad
J_y \le J_z + J_x, \qquad
J_z \le J_x + J_y ,
\end{equation}
and in one of the gapped $A$ phases when one of them is violated,
\begin{align}
A_x &: \; J_x > J_y + J_z, \\
A_y &: \; J_y > J_z + J_x, \\
A_z &: \; J_z > J_x + J_y .
\end{align}
The phase transitions occur when one of the triangle inequalities is
saturated. Accordingly, the phase boundaries between the gapless $B$
phase and the three gapped phases are given by
\begin{equation}
\begin{aligned}
A_x\!-\!B &: \quad J_x = J_y + J_z,\\
A_y\!-\!B &: \quad J_y = J_z + J_x,\\
A_z\!-\!B &: \quad J_z = J_x + J_y .
\end{aligned}
\label{eq:phase_boundaries}
\end{equation}

\begin{figure}[!tbp]
\centering
\includegraphics[width=0.92\columnwidth]{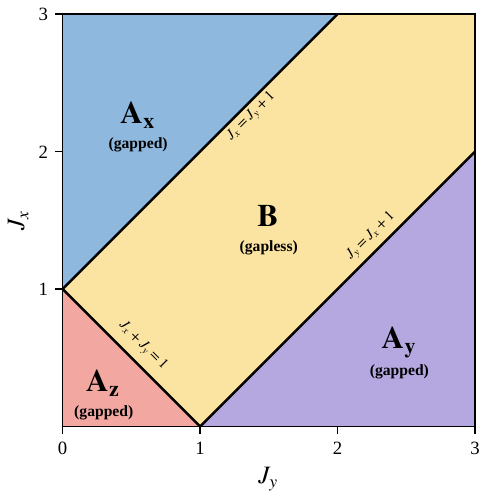}
\caption{Phase diagram of the Kitaev model in the $(J_x,J_y)$ plane at
$J_z=1$ (schematic).}
\label{fig:phases}
\end{figure}

The resulting phase diagram is shown schematically in Fig.~\ref{fig:phases}.
In the present convention, we fix the overall energy scale by setting
$J_z=1$. The remaining parameter space is therefore represented in the
$(J_x,J_y)$ plane. The phase boundaries in this plane follow directly
from Eq.~(\ref{eq:phase_boundaries}) upon setting $J_z=1$. These
boundaries divide the parameter space into the central gapless $B$ phase
and the three gapped phases $A_x$, $A_y$, and $A_z$, corresponding to the
regions in which $J_x$, $J_y$, or $J_z$, respectively, is the dominant
coupling.

\section{Ground State of the Kitaev Honeycomb Model}
\label{sec:gs}

In this section, to calculate the quantum geometric tensor and the associated geometric
quantities, which provide valuable information about the geometry of the
ground-state manifold and the phase transitions of the model, we first
investigate the ground state of the Kitaev model described by the Hamiltonian
in Eq.~(\ref{HKM}).

The Kitaev model can be solved exactly by representing each spin in terms of
four Majorana fermions,
\begin{equation}
\sigma_i^\alpha = i\,b_i^\alpha c_i ,
\end{equation}
where $c_i$ denotes the itinerant Majorana fermion, while $b_i^\alpha$,
with $\alpha=x,y,z$, are the gauge Majorana operators. The physical Hilbert
space is selected by imposing the local constraint
$D_i=b_i^xb_i^yb_i^zc_i=1$ \cite{kitaev2006anyons}.

In the Majorana representation, the Hamiltonian takes the quadratic form
\begin{equation}
H = \frac{i}{4}\sum_{j,k} A_{jk}\,c_jc_k.
\end{equation}
Here, $A_{jk}=2J_{\alpha_{jk}}\,u_{jk}$ is the coupling matrix, where
$J_{\alpha_{jk}}$ denotes the exchange coupling associated with the link
connecting sites $j$ and $k$. The corresponding link variable is
\begin{equation}
u_{jk}=i\,b_j^{\alpha_{jk}}b_k^{\alpha_{jk}},
\end{equation}
with $\alpha_{jk}$ specifying the type of that link. The link variables
$u_{jk}=\pm1$ commute with the Hamiltonian. The plaquette
operators are also conserved quantities and are defined as
\begin{equation}
W_p=
\sigma^x_1\sigma^y_2\sigma^z_3
\sigma^x_4\sigma^y_5\sigma^z_6
=
\prod_{(j,k)\in\partial p}u_{jk},
\qquad W_p^2=1.
\end{equation}

According to Lieb's theorem~\cite{lieb1994flux}, the ground state lies in the vortex-free sector,
characterized by $W_p=+1$ for every plaquette. In this sector, a uniform gauge
can be chosen, restoring translational invariance. Consequently, the original
spin problem reduces to free Majorana fermions hopping on the honeycomb
lattice, which has a two-site unit cell. Since a global rescaling of the couplings,
$J_\alpha\to cJ_\alpha$ with $c>0$, does not change the ground state, only
the ratios of the exchange couplings are physically relevant. We therefore
fix the overall energy scale by setting $J_z=1$. The ground-state manifold is then two-dimensional and parametrized by
$\lambda=(\lambda^1,\lambda^2)=(J_x,J_y)$. These two parameters define the coordinates of the ground-state
manifold considered in the subsequent geometric analysis.
To determine the ground-state wave function explicitly, we now
express the Hamiltonian in momentum space by introducing the
Fourier transforms of the Majorana operators on the two
sublattices of the honeycomb lattice.

Let $a_{q}$ and $b_{q}$ denote the Fourier transforms of
the Majorana operators $c_i$ on the two sublattices of the honeycomb lattice.
Because the Majorana operators satisfy
$a_{-q}=a^\dagger_{q}$ and
$b_{-q}=b^\dagger_{q}$, the momentum modes
$q$ and $-q$ are not independent. The Hamiltonian can
therefore be decomposed into independent two-level problems over half of
the Brillouin zone (BZ):
\begin{equation}
H=
\sum_{q\in{\rm BZ}/2}
\Psi^\dagger_{q}\,H(q)\,\Psi_{q},
\end{equation}
where
$\Psi_{q}=(a_{q},\, b_{q})^{T}$
is the two-component momentum-space spinor, and $H(q)$ is the
corresponding Bloch Hamiltonian.

After a $q$- and $\lambda$-independent phase redefinition of one
sublattice, the Bloch Hamiltonian takes the two-band form
\begin{equation}
H(q)=
\begin{pmatrix}
0 & f(q)\\
f^*(q) & 0
\end{pmatrix}.
\end{equation}
The off-diagonal function is given by
\begin{equation}\label{fq}
f(q)=2\left(J_x e^{iq_1}+J_y e^{iq_2}+1\right),
\end{equation}
where $q_1=q\cdot n_1$ and $q_2=q\cdot n_2$.
Here, $n_1$ and $n_2$ are the primitive lattice vectors of
the honeycomb lattice, and the constant term corresponds to the choice
$J_z=1$.

The energy spectrum consists of two bands,
\begin{equation}
E_\pm(q)=\pm\epsilon(q),
\qquad
\epsilon(q)=|f(q)|.
\end{equation}
Thus, the positive and negative energy bands are determined by the
magnitude of the complex function $f(q)$.

Having obtained the energy spectrum, we now determine the
ground-state wave function by finding the normalized eigenstate
associated with the negative-energy band for each momentum mode.
To this end, we write the
off-diagonal function in polar form as
\begin{equation}
f(q)=|f(q)|e^{i\varphi_q},
\end{equation}
where
$\varphi_q$ is its phase. The normalized eigenstate associated
with the negative-energy band is then
\begin{equation}
|u_-(q)\rangle=
\frac{1}{\sqrt2}
\begin{pmatrix}
-e^{i\varphi_{q}}\\
1
\end{pmatrix}.
\end{equation}
The many-body ground state is obtained by occupying the negative-energy
band for every independent momentum mode in half of the Brillouin zone.
It can therefore be expressed as
\begin{equation}\label{si0}
|\psi_0\rangle=
\bigotimes_{q\in\mathrm{BZ}/2}
|u_-(q)\rangle.
\end{equation}

\section{Quantum Geometry of the Ground State}\label{sec:qgt}

In this section, we investigate the geometric tensor of the ground state
of the Kitaev model. Since $|\psi_0\rangle$ is a product over modes, its
geometric tensor is the sum of the single-mode tensors,
\begin{equation}
\chi_{\mu\nu}
=
\sum_{q\in\mathrm{BZ}/2}
\chi_{\mu\nu}(q).
\end{equation}
The single-mode geometric tensor $\chi_{\mu\nu}(q)$ is given by
\begin{equation}
\chi_{\mu\nu}(q)
=
\langle\partial_\mu u_-|
\left(1-|u_-\rangle\langle u_-|\right)
|\partial_\nu u_-\rangle.
\end{equation}
For the Kitaev model, the geometric tensor takes the form
\begin{equation}
\chi_{\mu\nu}(q)
=
\frac14\,
\partial_\mu\varphi_q\,
\partial_\nu\varphi_q,
\end{equation}
which is real and symmetric. Consequently, the Berry curvature
$F_{\mu\nu}=-2\,\mathrm{Im}[\chi_{\mu\nu}]$ vanishes identically on the
$(J_x,J_y)$ manifold, and the geometric tensor coincides with the quantum
metric,
\begin{equation}\label{giD}
g_{\mu\nu}
=
\frac14
\sum_{q\in\mathrm{BZ}/2}
\partial_\mu\varphi_q\,
\partial_\nu\varphi_q .
\end{equation}

In the thermodynamic limit, the sum over momentum modes is replaced by an
integral according to
\begin{equation}
\sum_{q\in\mathrm{BZ}/2}
\longrightarrow
\frac{N}{2}
\int_{\mathrm{BZ}}\frac{d^2q}{(2\pi)^2},
\end{equation}
where $N$ is the number of unit cells and the integration variables are
$(q_1,q_2)\in[0,2\pi)^2$, so that the Brillouin-zone measure is
normalized to unity. Since $\varphi_{-q}=-\varphi_q$, the integrand is even in
$q$ and the integral may be taken over the full BZ. Consequently, the
intensive quantum metric, defined per unit cell, becomes
\begin{equation}\label{gN}
\frac{g_{\mu\nu}}{N}
=
\frac18
\int_{\mathrm{BZ}}\frac{d^2q}{(2\pi)^2}\,
\partial_\mu\varphi_q\,
\partial_\nu\varphi_q.
\end{equation}
The derivatives of the phase $\varphi_q$ with respect to the model
parameters are expressed in terms of the complex function $f(q)$ as
\begin{equation}\label{dphi}
\partial_\mu\varphi_q
=
\mathrm{Im}\!\left[
\frac{\partial_\mu f(q)}{f(q)}
\right]
=
\mathrm{Im}\!\left[
\frac{e^{iq_\mu}}{1+J_xe^{iq_1}+J_ye^{iq_2}}
\right],
\end{equation}
where $\mu=x,y$ is identified with $q_\mu=q_1,q_2$, respectively.

Throughout this paper, all curvatures are
computed from the intensive metric $g_{\mu\nu}/N$ of Eq.~(\ref{gN}). Since the
scalar curvature transforms as
\begin{equation}
g_{\mu\nu}\to c\,g_{\mu\nu}\quad\Longrightarrow\quad R\to R/c,
\label{Rscale}
\end{equation}
the scalar curvature of the extensive metric $g_{\mu\nu}$ is smaller by a
factor $N$. Other common conventions, such as defining the distance from the
fidelity itself rather than from its square as in Eq.~(\ref{DS2}), halve the
metric and therefore rescale $R$ by a constant factor.
The signs, the locations of the divergences, and the
critical exponents are independent of this choice.

\section{Geometry of the Gapped Phases}
\label{sec:gapped}

In this section, we calculate the metric and the Ricci scalar in the gapped
phases of the Kitaev honeycomb model and derive their analytical
expressions in terms of the coupling parameters.

\subsection{\texorpdfstring{Phase $A_z$}{Phase Az}}

In the gapped phase $A_z$, defined by $J_x+J_y<1$, the modulus
of $J_xe^{iq_1}+J_ye^{iq_2}$ is smaller than one for all $q$. The function
appearing in Eq.~(\ref{dphi}) can then be expanded in a geometric series,
\begin{equation}
\frac{1}{1+J_xe^{iq_1}+J_ye^{iq_2}}=\sum_{n=0}^{\infty}(-1)^n\left(J_xe^{iq_1}+J_ye^{iq_2}\right)^n,
\end{equation}
which contains only non-negative Fourier harmonics. We write the integrand of
Eq.~(\ref{gN}) with the identity
\begin{equation}
\mathrm{Im}\,w_\mu\,\mathrm{Im}\,w_\nu=\tfrac12\left[\mathrm{Re}(w_\mu\bar w_\nu)-\mathrm{Re}(w_\mu w_\nu)\right],
\end{equation}
where $\mu,\nu=1,2$ and $w_\mu$ is given by
\begin{equation}
w_\mu=\frac{e^{iq_\mu}}{1+J_xe^{iq_1}+J_ye^{iq_2}}.
\end{equation}
The products
$w_\mu w_\nu$ contain only strictly positive harmonics and integrate to zero,
and the remaining integrals are elementary. We obtain
\begin{align}
\frac{g_{xx}}{N}
=
\frac{g_{yy}}{N}
&=
\frac{1}{16}\int_{\mathrm{BZ}}\frac{d^2q}{(2\pi)^2}
\frac{1}{|1+J_xe^{iq_1}+J_ye^{iq_2}|^2}
\nonumber\\
&=
\frac{1}{16\sqrt{D}},
\label{gxx_gyy_Az}
\\[4pt]
\frac{g_{xy}}{N}
&=
\frac{1-J_x^2-J_y^2-\sqrt{D}}
{32J_xJ_y\sqrt{D}},
\label{gxy_Az}
\end{align}
where
\begin{equation}
D=
\left[1-(J_x+J_y)^2\right]
\left[1-(J_x-J_y)^2\right].
\label{D_Az}
\end{equation}
Within the $A_z$ phase $D>0$, whereas $D\to0$ at the phase boundary.

The equality $g_{xx}=g_{yy}$ in Eq.~(\ref{gxx_gyy_Az}) holds pointwise in
the $A_z$ phase, but it is not a consequence of the exchange symmetry
of the model. The symmetry $J_x\leftrightarrow J_y$, together with
$q_1\leftrightarrow q_2$, only implies
\begin{equation}
g_{xx}(J_x,J_y)=g_{yy}(J_y,J_x).
\end{equation}
The equality at the same point is a special property of the $A_z$ phase, in
which both diagonal components reduce to the same integral in
Eq.~(\ref{gxx_gyy_Az}); as shown below, it does not hold in the $A_x$ and
$A_y$ phases.

Using these metric components, the Christoffel symbols, the Riemann tensor,
the Ricci tensor, and the scalar curvature follow from
Eqs.~(\ref{Christoffel})-(\ref{Ricci}). The resulting Ricci scalar of the
ground-state manifold in the $A_z$ phase is
\begin{equation}
R_{A_z}
=
-16\,
\frac{1-J_x^2-J_y^2+\sqrt{D}}{D},
\qquad J_x+J_y<1.
\label{RAz}
\end{equation}

\subsection{\texorpdfstring{Phases $A_x$ and $A_y$: exact mapping onto $A_z$}{Phases Ax and Ay: exact mapping onto Az}}

The metric in the phases $A_x$ and $A_y$ cannot be obtained by a simple
permutation of $J_x$, $J_y$, and $J_z$ in Eqs.~(\ref{gxx_gyy_Az}) and
(\ref{gxy_Az}), because the coordinates $(J_x,J_y)$ are defined at fixed
$J_z=1$ and are not permuted together with the couplings. Instead, we use the
fact that only coupling ratios matter. In the $A_x$ phase we factor out the
dominant term,
\begin{equation}
f(q)=2J_xe^{iq_1}\left[1+u\,e^{ip_1}+v\,e^{ip_2}\right],
\label{mapAx}
\end{equation}
with
\begin{equation}
u=\frac{1}{J_x},\qquad v=\frac{J_y}{J_x},\qquad p_1=-q_1,\qquad p_2=q_2-q_1 .
\label{uv}
\end{equation}
The prefactor contributes the coupling-independent phase $q_1$, and the change
of variables $(q_1,q_2)\mapsto(p_1,p_2)$ is a measure-preserving bijection of
the BZ. Hence
\begin{equation}
\varphi_q=q_1+\varphi^{(A_z)}_p(u,v),
\end{equation}
where $\varphi^{(A_z)}$ is the phase of the $A_z$ problem with couplings
$(u,v)$, and the condition $J_x-J_y>1$ is equivalent to $u+v<1$. The metric of
the $A_x$ phase is therefore the pullback of the $A_z$ metric under the map
of Eq.~(\ref{uv}),
\begin{equation}
g^{(A_x)}_{\mu\nu}(J_x,J_y)=\frac{\partial u^a}{\partial\lambda^\mu}\frac{\partial u^b}{\partial\lambda^\nu}\,
g^{(A_z)}_{ab}(u,v).
\end{equation}
Using the relation
\begin{equation}
D(u,v)=\frac{D(J_x,J_y)}{J_x^4},
\end{equation}
we obtain explicitly
\begin{align}
\frac{g_{xx}}{N}&=\frac{J_x^2-\sqrt{D}}{16J_x^2\sqrt{D}}, \qquad
\frac{g_{yy}}{N}=\frac{1}{16\sqrt{D}},
\label{gAx_diag}\\[4pt]
\frac{g_{xy}}{N}&=-\,\frac{J_x^2+J_y^2-1-\sqrt{D}}{32J_xJ_y\sqrt{D}},
\qquad J_x-J_y>1,
\label{gAx_off}
\end{align}
with the same function $D$ as in Eq.~(\ref{D_Az}), which is positive
throughout $A_x$. Note that $g_{xx}\neq g_{yy}$ in this phase. The metric of
the $A_y$ phase follows from Eqs.~(\ref{gAx_diag}) and (\ref{gAx_off}) by
the exchange $x\leftrightarrow y$:
\begin{align}
\frac{g_{yy}}{N}&=\frac{J_y^2-\sqrt{D}}{16J_y^2\sqrt{D}}, \qquad
\frac{g_{xx}}{N}=\frac{1}{16\sqrt{D}},\\[4pt]
\frac{g_{xy}}{N}&=-\,\frac{J_x^2+J_y^2-1-\sqrt{D}}{32J_xJ_y\sqrt{D}},
\qquad J_y-J_x>1.
\end{align}
Since the scalar curvature is a coordinate invariant,
\begin{equation}
R_{A_x}(J_x,J_y)=R_{A_z}\!\left(\frac{1}{J_x},\frac{J_y}{J_x}\right),
\end{equation}
which gives
\begin{align}
R_{A_x}
&=
-16\,J_x^2\,
\frac{J_x^2-J_y^2-1+\sqrt{D}}{D},
\qquad J_x-J_y>1,
\label{RAx}
\\[4pt]
R_{A_y}
&=
-16\,J_y^2\,
\frac{J_y^2-J_x^2-1+\sqrt{D}}{D},
\qquad J_y-J_x>1.
\label{RAy}
\end{align}
The different functional forms of Eqs.~(\ref{RAz}), (\ref{RAx}), and
(\ref{RAy}) therefore reflect only the choice of coordinates: intrinsically,
the three gapped phases have the same geometry.

\section{Geometry of the Gapless Phase}
\label{sec:gapless}

In this section, we examine the ground state of the Kitaev honeycomb model in
the gapless phase and calculate the corresponding Ricci scalar, focusing on
its dependence on the coupling parameters and the system size. In phase $B$,
the ground state is the product state of Eq.~(\ref{si0}), but, unlike in the
gapped phases, the Bloch function $f(q)$ vanishes at two Dirac points
$\pm q_D=\pm(q_1^*,q_2^*)$, where $\varphi_q$ is ill-defined and exhibits a
vortex structure. The condition $f(q_D)=0$ reads
\begin{equation}
1+J_x e^{iq_1^*}+J_y e^{iq_2^*}=0.
\label{Dirac}
\end{equation}
Geometrically, Eq.~(\ref{Dirac}) states that $1$, $J_x$, and $J_y$ form a
closed triangle in the complex plane, which is possible only when the
triangle inequalities hold, i.e., in phase $B$.

\subsection{Logarithmic divergence of the metric}

Near the Dirac points, $\partial_\mu\varphi_q\sim1/|q\mp q_D|$, so the
integral in Eq.~(\ref{gN}) diverges logarithmically: in the thermodynamic
limit the intensive metric of phase $B$ is infinite, and a finite-size
regularization is required. Linearizing $f(q)$ around $\pm q_D$ and cutting
the momentum integral at a distance $\epsilon$ from the Dirac points gives
(Appendix~\ref{app:dirac})
\begin{equation}
\begin{aligned}
\frac{g_{xx}}{N}
=
\frac{g_{yy}}{N}
&=
\frac{\ln(1/\epsilon)}{8\pi\sqrt{-D}}+O(1),\\
\frac{g_{xy}}{N}
&=
\frac{(1-J_x^2-J_y^2)\ln(1/\epsilon)}
{16\pi J_xJ_y\sqrt{-D}}+O(1),
\end{aligned}
\label{gB}
\end{equation}
where
\begin{equation}
D=
\left[1-(J_x+J_y)^2\right]
\left[1-(J_x-J_y)^2\right]<0
\end{equation}
in phase $B$. The coefficients of the logarithms are universal: they do not
depend on the shape of the excluded region or on the prefactor of $\epsilon$,
which only affect the $O(1)$ terms. For a system of $N=L^2$ unit cells the
natural infrared scale is the momentum spacing, $\epsilon\sim2\pi/L$, so
that $\ln(1/\epsilon)=\ln L+O(1)$.

\subsection{Infrared regularization}

On a finite lattice with periodic boundary conditions, the metric is a
discrete sum over the allowed momenta $(q_1,q_2)=2\pi(m_1,m_2)/L$ rather than a
cut-off integral, and the identification $\epsilon\sim2\pi/L$ requires care. The
Dirac points generically lie at a distance $\delta_{\min}<\sqrt2\pi/L$ from the
nearest allowed momentum, and this single mode contributes a term of order
$1/(L\delta_{\min})^2$ to $g_{\mu\nu}/N$. This term is unbounded: when a Dirac
point coincides with an allowed momentum, the finite-size ground state is
degenerate and the metric diverges. The lattice metric therefore shows sharp
resonances as a function of $L$ and of the couplings, consistent with the
peaks of the fidelity susceptibility observed in Ref.~\cite{yang2008fidelity}.
Away from the resonances, we find that it grows as $\ln L$ with the slope of
Eq.~(\ref{gB}). For the curvature the resonances are even more
severe, because the position of $q_D$ relative to the momentum grid changes
with the couplings on the scale $\Delta J\sim1/L$, so that the second
derivatives of the resonant term grow as $L^2$. The curvature of the raw
lattice metric is therefore not representative of the macroscopic geometry of
the manifold.

We thus define the curvature of phase $B$ through the smooth part of the
metric, $g^{(L)}_{\mu\nu}$, obtained from the integral of Eq.~(\ref{gN}) with
two disks of radius $\epsilon_L=2\pi/L$ around $\pm q_D$ excluded in the
$(q_1,q_2)$ plane. This metric is a smooth function of the couplings,
reproduces the $\ln L$ growth of the non-resonant lattice sums, and differs
from them only in the non-universal $O(1)$ terms, which affect only the
subleading corrections to the curvature discussed below.

\subsection{Ricci scalar}

Keeping the leading logarithmic terms of Eq.~(\ref{gB}) with
$\ln(1/\epsilon)\to\ln L$, the metric is nondegenerate,
\begin{equation}
\det\!\left(\frac{g_{\mu\nu}}{N}\right)=\frac{\ln^2L}{256\pi^2J_x^2J_y^2}.
\end{equation}
Computing the Christoffel symbols, the Riemann and Ricci tensors, and then the
scalar curvature from Eqs.~(\ref{Christoffel})-(\ref{Ricci}), the Ricci scalar
of phase $B$ is obtained as
\begin{equation}
R_B=
\frac{\bar{R}_B}{\ln L}\left[1+O\!\left(\frac{1}{\ln L}\right)\right],
\label{RB}
\end{equation}
where
\begin{equation}
\bar{R}_B=\frac{128\pi J_x^2J_y^2}{(-D)^{3/2}},
\label{RBbar}
\end{equation}
valid for $|J_x-J_y|<1<J_x+J_y$. For a metric of the form
$g_{\mu\nu}=A_{\mu\nu}\ln L+B_{\mu\nu}$, Eq.~(\ref{Rscale}) implies
\begin{equation}
R[g]=\frac{R[A]}{\ln L}+O\!\left(\frac{1}{\ln^2L}\right),
\end{equation}
so the $O(1)$ terms of the metric contribute only to the relative correction
of order $1/\ln L$ in Eq.~(\ref{RB}). These corrections are numerically
significant for realistic sizes, and they grow close to the phase boundaries;
the ratio of $R_B$ to $\bar{R}_B/\ln L$ approaches unity as $1/\ln L$. We have
also checked that the curvature of the regularized metric is positive at
representative points throughout phase $B$, including points close to all
three critical lines.

Three features of Eq.~(\ref{RB}) are important. First, the curvature of the
smooth metric is positive throughout phase $B$, in contrast to the gapped
phases. Second, $R_B\to0^+$ as $L\to\infty$: in the thermodynamic limit the
metric diverges and the manifold becomes flat on the scale of the intensive
metric. The meaningful $L$-independent quantity characterizing the gapless
phase is therefore the amplitude
\begin{equation}
\bar{R}_B=\lim_{L\to\infty}R_B\ln L .
\end{equation}
Third, $\bar{R}_B$ diverges at the phase boundaries. Near the $A_z$-$B$ line,
where $\delta=1-J_x-J_y\to0^-$ and $-D\simeq8J_xJ_y|\delta|$,
\begin{equation}
\bar{R}_B\simeq\frac{4\sqrt2\,\pi\sqrt{J_xJ_y}}{|\delta|^{3/2}},
\label{RB_crit}
\end{equation}
so that the exponent of the divergence ($3/2$) differs from that of the gapped
side ($1$), Eq.~(\ref{RAz_crit}). Equation~(\ref{RB}) requires the two Dirac
points to be resolved by the infrared cutoff. Since they merge at the phase
boundary with a separation proportional to $|\delta|^{1/2}$, this condition
breaks down in a crossover region $|\delta|\lesssim C\,(2\pi/L)^2$, which
shrinks rapidly with increasing $L$. The constant $C$ is not small: inside this
region the curvature of the regularized metric can even become negative. There,
neither Eq.~(\ref{RB}) nor the regularization is reliable, and the
finite-size geometry must be studied separately.

\section{Phase Diagram from the Ricci Scalar}
\label{sec:map}

In this section, we investigate the phase transitions of the Kitaev honeycomb
model using the Ricci scalar approach. We demonstrate how the behavior of the
scalar curvature can be used to identify the phase boundaries and reconstruct
the phase diagram of the model. Having obtained the scalar curvature of the
ground-state manifold in all four phases, as given by Eqs.~(\ref{RAz}),
(\ref{RAx}), (\ref{RAy}), and (\ref{RB}), we now examine how the Ricci scalar
characterizes the phase structure of the Kitaev model. As discussed in
Sec.~\ref{sec:model} and illustrated in Fig.~\ref{fig:phases}, the phase
diagram is determined by the relative magnitudes of the exchange couplings.
Setting $J_z=1$, the gapless phase $B$ is bounded by the critical lines
$J_x+J_y=1$ and $|J_x-J_y|=1$. At these phase boundaries, the excitation gap
closes and the quantity $D$ of Eq.~(\ref{D_Az}) vanishes. The scalar curvature
therefore provides a geometric probe of the phase transitions through its
behavior near these critical lines. We restrict the analysis to
$J_x,J_y\geq0$, because the signs of the couplings can be absorbed into
momentum shifts and do not affect the scalar curvature.

\begin{figure}[!tbp]
\centering
\includegraphics[width=\columnwidth]{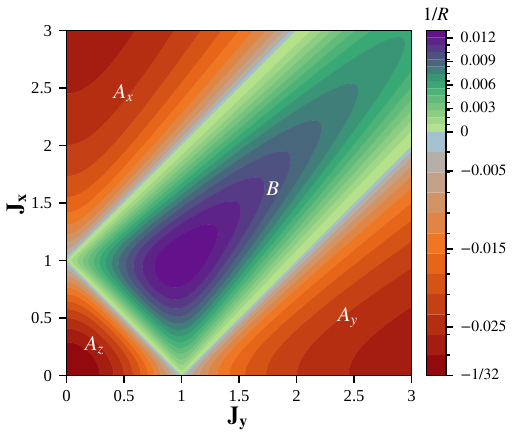}\\[-2pt]
{\small (a)}\\[6pt]
\includegraphics[width=\columnwidth]{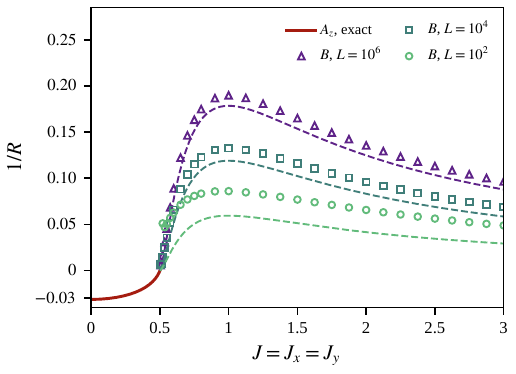}\\[-2pt]
{\small (b)}
\caption{Inverse scalar curvature at zero field and $J_z=1$. (a) $1/R$ in the
$(J_x,J_y)$ plane, with $1/\bar{R}_B$ in phase $B$. (b) $1/R$ along
$J_x=J_y=J$; in phase $B$ for $L=10^2$, $10^4$, and $10^6$ (symbols) together
with $\ln L/\bar{R}_B$ (dashed lines).}
\label{fig:map}
\end{figure}

To visualize the phase structure, we plot inverse curvatures, which remain
finite and vanish where the curvature diverges. In the gapped phases the
curvature is exact in the thermodynamic limit, and we plot $1/R$. In phase $B$,
$R_B$ itself vanishes as $1/\ln L$, so a map of $1/R_B$ would depend on an
arbitrary choice of $L$. We therefore plot the $L$-independent quantity
$1/\bar{R}_B$, which determines the leading behavior of the inverse curvature,
\begin{equation}
\frac{1}{R_B}\simeq\frac{\ln L}{\bar{R}_B},
\end{equation}
up to corrections of relative order $1/\ln L$. With this choice, no assumption
on the system size enters the phase map. For phase $B$ we thus take
$1/\bar{R}_B$ as the counterpart of $1/R$ in order to display the phase
transition. The resulting map in the $(J_x,J_y)$ plane is shown in panel (a)
of Fig.~\ref{fig:map}; both $1/R$ and $1/\bar{R}_B$ vanish on the critical
lines.

The resulting map reproduces the four-phase structure shown in
Fig.~\ref{fig:phases}: the gapped phase $A_z$ lies near the origin, the
phases $A_x$ and $A_y$ occupy the regions on either side of the gapless phase,
and phase $B$ lies between the critical lines. In the gapped phases the Ricci
scalar is negative, whereas it is positive in phase $B$. Accordingly, the
inverse curvature changes sign across the phase boundaries and approaches zero
where the curvature diverges.

Panel (b) of Fig.~\ref{fig:map} shows a cut along the diagonal $J_x=J_y=J$, which
crosses the $A_z$-$B$ transition at $J=1/2$. In the $A_z$ phase the solid line
is the exact $1/R$. In phase $B$ the symbols show $1/R$ of the regularized
metric for different values of $L$, and the dashed
lines the leading behavior $\ln L/\bar{R}_B$ for the same sizes; points in the crossover region,
where the separation of the Dirac points is smaller than $12\times2\pi/L$, are
omitted. The finite-size dependence of phase $B$ is thus made explicit: the
inverse curvature of the regularized metric is positive, increases as $\ln L$,
and approaches the asymptotic form with corrections of relative order
$1/\ln L$. On both
sides of the transition the inverse curvature vanishes at $J=1/2$, with
the different exponents of Eqs.~(\ref{RAz_crit}) and (\ref{RB_crit}).

\section{Effect of a Magnetic Field}
\label{sec:field}

\subsection{Model and quantum metric}

A weak magnetic field $\mathbf{h}$ breaks time-reversal symmetry. To third
order in perturbation theory it generates a three-spin interaction which
preserves the exact solvability of the model~\cite{kitaev2006anyons},
\begin{equation}
H_K=-K\sum_{(j,k,l)}\sigma^x_j\sigma^y_k\sigma^z_l,
\qquad K\sim\frac{h_xh_yh_z}{J^2},
\label{HK}
\end{equation}
where the sum runs over the triples of sites specified in
Ref.~\cite{kitaev2006anyons}. In the vortex-free sector, $H_K$ adds a
next-nearest-neighbor hopping of the Majorana fermions, which appears as a
mass term on the diagonal of the Bloch Hamiltonian,
\begin{equation}
H(q)=
\begin{pmatrix}
\Delta_q & f(q)\\
f^*(q) & -\Delta_q
\end{pmatrix},
\label{HqK}
\end{equation}
where $\Delta_q$ is defined as
\begin{equation}
\Delta_q=4K\left[\sin q_1-\sin q_2+\sin(q_2-q_1)\right],
\label{Deltaq}
\end{equation}
with $f(q)$ given by Eq.~(\ref{fq}) and $K$ measured in units of $J_z=1$. The
spectrum is
\begin{equation}
E_\pm(q)=\pm\sqrt{|f(q)|^2+\Delta_q^2}.
\end{equation}
The mass $\Delta_q$
vanishes at the points $(\pi,\pi)$, $(\pi,0)$, and $(0,\pi)$ at which the two
Dirac points merge on the $A_z$-$B$, $A_x$-$B$, and $A_y$-$B$ lines,
respectively. Inside phase $B$, however, $\Delta_{\pm q_D}\neq0$, so this phase
becomes gapped and acquires a Chern number $\nu=\pm1$; this is the
non-Abelian phase of the model, while the $A$ phases remain Abelian with
$\nu=0$~\cite{kitaev2006anyons}.

For each momentum mode, Eq.~(\ref{HqK}) describes a spin-$1/2$ in the
effective field
\begin{equation}
\mathbf{b}_q=\left(\mathrm{Re}f(q),\,-\mathrm{Im}f(q),\,\Delta_q\right),
\qquad
\hat{\mathbf{n}}_q=\frac{\mathbf{b}_q}{|\mathbf{b}_q|},
\end{equation}
and the ground state of the mode is determined by the unit vector
$\hat{\mathbf{n}}_q$. The single-mode quantum metric is
\begin{equation}
g_{\mu\nu}(q)
=\frac14\,\partial_\mu\hat{\mathbf{n}}_q
\cdot\partial_\nu\hat{\mathbf{n}}_q,
\end{equation}
and the intensive metric generalizing Eq.~(\ref{gN}) reads
\begin{equation}
\frac{g_{\mu\nu}}{N}
=\frac18\int_{\mathrm{BZ}}\frac{d^2q}{(2\pi)^2}\,
\frac{\partial_\mu\mathbf{b}_q\cdot\partial_\nu\mathbf{b}_q
-(\hat{\mathbf{n}}_q\cdot\partial_\mu\mathbf{b}_q)(\hat{\mathbf{n}}_q\cdot\partial_\nu\mathbf{b}_q)}
{|\mathbf{b}_q|^2},
\label{gNK}
\end{equation}
which reduces to Eq.~(\ref{gN}) for $K=0$. Since $\hat{\mathbf n}_q$ now leaves
the equatorial plane, the Berry curvature on the $(J_x,J_y)$ manifold no longer
vanishes; as before, we focus on the metric. The field adds a third parameter
to the ground-state manifold. We keep $K$ fixed (in units of $J_z$) and study
the two-dimensional manifold of couplings $(J_x,J_y)$ with $J_z=1$, as in the
previous sections. Note that the scalar curvature depends on this choice of
slice: a different two-dimensional section of the three-dimensional parameter
space has, in general, a different intrinsic curvature.

The metric of Eq.~(\ref{gNK}) depends on $K$ only through $\Delta_q^2$. In the
gapped phases, where $|\mathbf b_q|$ is bounded away from zero, it is an
analytic function of $K^2$, and so is the curvature, which therefore reduces
smoothly to Eqs.~(\ref{RAz})-(\ref{RAy}) as $K\to0$. No closed form is
available for $K\neq0$, and we evaluate the metric and its first and second
derivatives with respect to $(J_x,J_y)$ by differentiating the integrand of
Eq.~(\ref{gNK}) analytically and integrating numerically, with the quadrature
refined around the points where the gap is
smallest. The scalar curvature then follows without numerical differentiation.
For $K\to0$ this procedure reproduces Eqs.~(\ref{RAz})-(\ref{RAy}) to
better than $10^{-8}$.

The quantum geometric tensor of the Kitaev model with the three-spin term was
studied by Lu and Wang~\cite{lu2025phase}, who computed the Berry curvature and
the Fubini-Study metric with the components of the effective field
$\mathbf b_q$ as parameters, and found that the metric reflects the phase
diagram, with peaks where critical lines cross. They did not compute the
curvature of the metric. Their analysis and ours agree on the points that can
be compared directly: the critical lines are independent of $K$, and the
field opens a gap in phase $B$ while leaving the $A$ phases gapped. In
Ref.~\cite{lu2025phase} the mass term is written as
\begin{equation}
\Delta^{\rm LW}_k=4K\left(\sin k_x-\sin k_y\right),
\end{equation}
without the third harmonic of Eq.~(\ref{Deltaq}) obtained by
Kitaev~\cite{kitaev2006anyons}. Both forms vanish at the merging
points of the Dirac cones, so the phase boundaries are the same.
All results below use Kitaev's form, Eq.~(\ref{Deltaq}).

\subsection{Phase diagram at finite field}\label{sec:finiteB}

\begin{figure}[!tbp]
\centering
\includegraphics[width=\columnwidth]{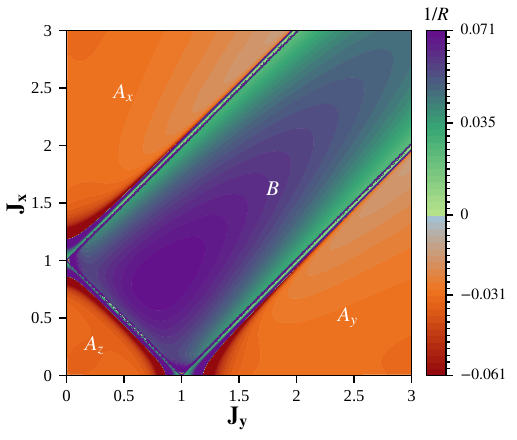}\\[-2pt]
{\small (a)}\\[6pt]
\includegraphics[width=\columnwidth]{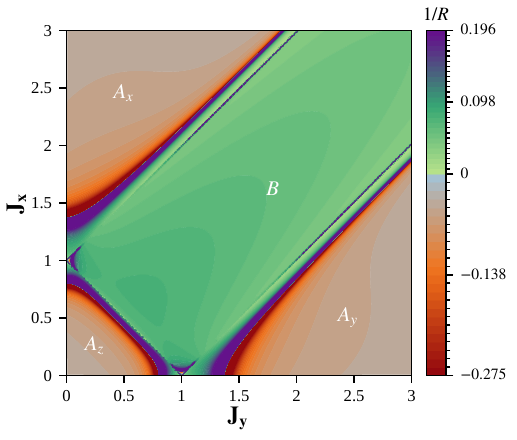}\\[-2pt]
{\small (b)}
\caption{Inverse scalar curvature $1/R$ in the $(J_x,J_y)$ plane at $J_z=1$
in a magnetic field: (a) $K=0.05$; (b) $K=0.1$.}
\label{fig:fieldmap}
\end{figure}

Figure~\ref{fig:fieldmap} shows the inverse scalar curvature $1/R$ in the
$(J_x,J_y)$ plane for $K=0.05$ and $K=0.1$. Away from the critical lines, the
map retains the structure of the zero-field map of Fig.~\ref{fig:map}: the
curvature is negative in the three Abelian phases and positive in the
non-Abelian phase $B$. The field reduces the magnitude of the negative
curvature of the $A$ phases, most strongly close to phase $B$. In phase $B$
the curvature is now finite in the thermodynamic limit (see
Sec.~\ref{sec:fieldIR}). The critical lines, on which the gap closes, appear as
lines on which $|R|$ becomes large and $1/R$ vanishes.

\begin{figure}[!tbp]
\centering
\includegraphics[width=\columnwidth]{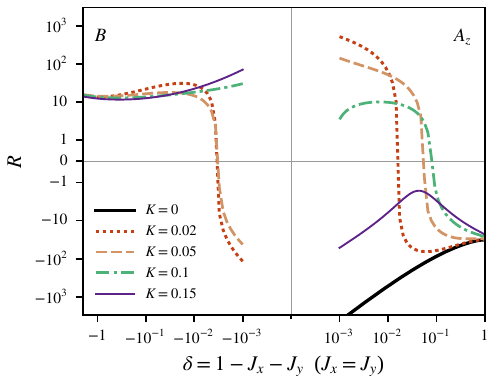}\\[-2pt]
{\small (a)}\\[6pt]
\includegraphics[width=\columnwidth]{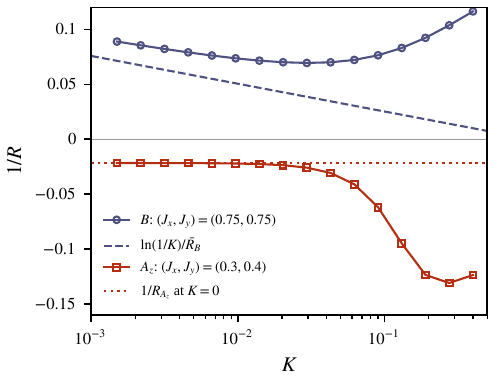}\\[-2pt]
{\small (b)}
\caption{(a) The scalar curvature $R$ as a function of $\delta=1-J_x-J_y$ along $J_x=J_y$ at
$J_z=1$, for $K=0$, $0.02$, $0.05$, $0.1$, and $0.15$. (b) The inverse scalar curvature $1/R$ as a function
of $K$ at $J_z=1$ for $(J_x,J_y)=(0.75,0.75)$ (phase $B$) and $(0.3,0.4)$
(phase $A_z$).}
\label{fig:fieldcut}
\end{figure}

The vicinity of the critical lines is more intricate than at $K=0$. Panel (a)
of Fig.~\ref{fig:fieldcut} shows the curvature along the diagonal $J_x=J_y$ as
a function of the distance $\delta=1-J_x-J_y$ from the $A_z$-$B$ critical
point, defined as in Sec.~\ref{sec:gapless}, for several values of $K$.
Positive (negative) $\delta$ corresponds to phase $A_z$ ($B$); at
$K=0$ only the $A_z$ side is shown, since $R_B\to0$ in the thermodynamic limit.
Both axes are logarithmic except for linear segments around zero
($|\delta|<10^{-3}$ and $|R|<1$), and the different values of $K$ are
distinguished by line style. Approaching the critical point from the $A_z$
side, the curvature first follows the zero-field curve, but in a crossover
region whose width grows as $K^2$ (Sec.~\ref{sec:scaling}) it deviates from it
and, for $K=0.02$ and
$K=0.05$, changes sign and becomes large and positive. On the $B$ side, the
curvature becomes large and negative for these values of $K$ in a narrow
region close to the critical point. For $K=0.1$ the curvature is positive on
both sides at distances $10^{-3}\lesssim|\delta|\lesssim3\times10^{-2}$, and
for $K=0.15$ the zero-field sign pattern holds at all distances we have
examined ($|\delta|\geq10^{-3}$), with $R<0$ on
the $A_z$ side and $R>0$ on the $B$ side and $|R|$ growing toward the critical
point. At $|\delta|=10^{-3}$, for example, we find
\begin{equation}\label{cross}
\left(R_{A_z},R_B\right)\approx
\begin{cases}
(516,\,-116), & K=0.02,\\
(141,\,-42), & K=0.05,\\
(3.5,\,30), & K=0.1,\\
(-52,\,70), & K=0.15.
\end{cases}
\end{equation}
In the maps of
Fig.~\ref{fig:fieldmap}, these crossover regions appear as narrow strips along
the critical lines in which $1/R$ takes large values of either sign. Thus, at
finite field the strong growth of $|R|$ still locates the critical lines, but
the sign of $R$ identifies the phases only outside a crossover region around
them. This behavior reflects the competition between the two small energy
scales near the transition, the distance $|\delta|$ from the critical line and
the field-induced mass $\Delta_q$, which vanishes linearly at the merging point
of the Dirac cones. A scaling description of this competition is given in
Sec.~\ref{sec:scaling}.

\subsection{\texorpdfstring{The field as an infrared cutoff in phase $B$}{The field as an infrared cutoff in phase B}}
\label{sec:fieldIR}

In phase $B$ the mass $\Delta_{\pm q_D}\propto K$ cuts off the logarithmic
divergence of the metric at momenta $|q\mp q_D|\sim K$, playing the role of
the infrared cutoff $2\pi/L$ of Sec.~\ref{sec:gapless}. The metric and the
curvature are therefore finite in the thermodynamic limit, and for $K\to0$ one
expects Eqs.~(\ref{gB}) and (\ref{RB}) with $\ln L$ replaced by $\ln(1/K)$:
\begin{equation}
\frac{g_{xx}}{N}\simeq\frac{\ln(1/K)}{8\pi\sqrt{-D}},
\qquad
\frac{1}{R_B(K)}\simeq\frac{\ln(1/K)}{\bar{R}_B}.
\label{RBK}
\end{equation}
We have verified the first relation numerically. As in the finite-size case,
the approach of the curvature to the second
relation is slow, with relative corrections of order $1/\ln(1/K)$. This is
shown in panel (b) of Fig.~\ref{fig:fieldcut} for $(J_x,J_y)=(0.75,0.75)$, where
the dashed line is the asymptotic form of Eq.~(\ref{RBK}): $R_B(K)$ is positive for all
$K$, is a nonmonotonic function of $K$ with a broad maximum
at intermediate $K$, and $1/R_B$ approaches the asymptotic form
$\ln(1/K)/\bar{R}_B$ from above as $K\to0$, with a slope that increases
toward $1/\bar{R}_B$. In the same panel, the curvature at a representative
point of the $A_z$ phase, $(J_x,J_y)=(0.3,0.4)$, approaches its zero-field
value $R_{A_z}$ of Eq.~(\ref{RAz}) (dotted line) quadratically in $K$ and decreases in magnitude
at larger $K$, while remaining negative.

\section{Scaling of the curvature near the critical line}
\label{sec:scaling}

The results of Sec.~\ref{sec:finiteB} show that the field opens a crossover
region around the critical lines, in which the sign of $R$ no longer identifies
the phase. Here we propose a scaling description of this region and test it
numerically. We restrict ourselves to the $A_z$-$B$ line along the diagonal
$J_x=J_y$ at $J_z=1$, and define $\delta=1-J_x-J_y$ as in
Sec.~\ref{sec:gapless}. Positive $\delta$ corresponds to the gapped phase
$A_z$, and negative $\delta$ to phase $B$.

We begin with the zero-field behavior. Near the $A_z$-$B$ line one has
$D\simeq8J_xJ_y\,\delta$ and $1-J_x^2-J_y^2\simeq2J_xJ_y$. Equation~(\ref{RAz})
then gives
\begin{equation}
R_{A_z}=-\frac{4}{\delta}\left[1+O(\delta^{1/2})\right],
\qquad \delta\to0^+,
\label{RAz_crit}
\end{equation}
away from the end points of the line. On the diagonal the two quantities above
reduce to $1-J_x^2-J_y^2\to1/2$ and $D\to2\delta$. In phase $B$ the
regularized curvature is $R_B=\bar R_B/\ln L$, where $\bar R_B$ diverges as in
Eq.~(\ref{RB_crit}). The exponents therefore differ on the two sides of the
transition. In the gapped phase the metric itself diverges algebraically,
$g/N\sim\delta^{-1/2}$ [Eq.~(\ref{gxx_gyy_Az})]. In phase $B$ the divergent
part is logarithmic in $L$, with a coefficient that diverges as
$|\delta|^{-1/2}$ [Eq.~(\ref{gB})].

To identify the relevant scale at finite field, we expand around the point
$(q_1,q_2)=(\pi,\pi)$ where the two Dirac points merge. We write
$q_i=\pi+\kappa_i$, with $s=\kappa_1+\kappa_2$ and $t=\kappa_1-\kappa_2$. On
the diagonal the Bloch function becomes
\begin{equation}
1+J_xe^{iq_1}+J_ye^{iq_2}\simeq\delta+\frac{J}{4}\,t^{2}-iJs ,
\end{equation}
so the Dirac points of phase $B$ sit at $t=\pm2\sqrt{|\delta|/J}$, and all
single-particle energies near the merging point are of order $|\delta|$. The
mass term of Eq.~(\ref{Deltaq}) becomes
\begin{equation}
\Delta_q\simeq-8K\,t .
\label{mass-merging}
\end{equation}
It vanishes linearly at the merging point and is of order $K|\delta|^{1/2}$ at
the Dirac points. The metric depends on $K$ only through $\Delta_q^2$
[Eq.~(\ref{gNK})], and $\Delta_q^2\sim K^2|\delta|$ must be compared with
$|f|^2\sim\delta^2$. The crossover is therefore controlled by
\begin{equation}
x=\frac{|\delta|}{K^{2}} ,
\label{x-def}
\end{equation}
and not by $|\delta|/K$. Since $R$ diverges as $1/\delta$ at $K=0$, we are led
to the scaling hypothesis
\begin{equation}
|\delta|\,R(K,\delta)\;\xrightarrow[K\to0]{}\;\mathcal F_\pm(x),
\label{scaling-form}
\end{equation}
with $\mathcal F_+$ for $\delta>0$ ($A_z$) and $\mathcal F_-$ for $\delta<0$
($B$). We have not derived Eq.~(\ref{scaling-form}) from the low-energy theory;
it is motivated by power counting, and we test it numerically.

\begin{figure*}[t]
\centering
\begin{minipage}[t]{0.32\textwidth}
  \centering
  \includegraphics[width=\linewidth]{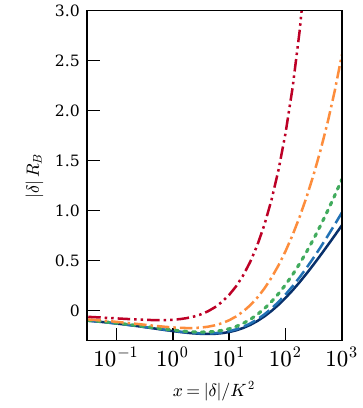}\\[2pt]
  {\small (a)}
\end{minipage}\hfill
\begin{minipage}[t]{0.32\textwidth}
  \centering
  \includegraphics[width=\linewidth]{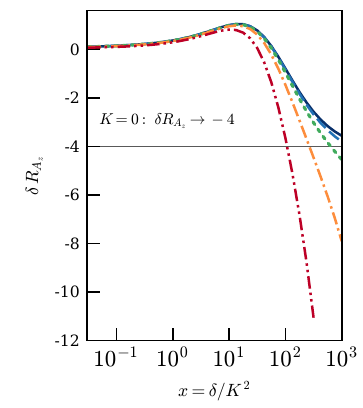}\\[2pt]
  {\small (b)}
\end{minipage}\hfill
\begin{minipage}[t]{0.32\textwidth}
  \centering
  \includegraphics[width=\linewidth]{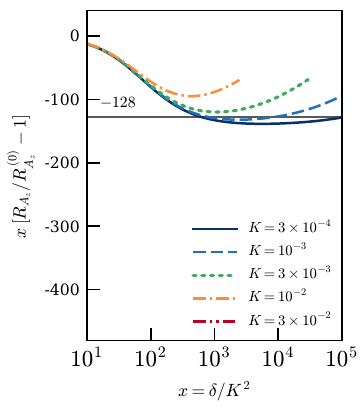}\\[2pt]
  {\small (c)}
\end{minipage}
\caption{Ricci scalar near the $A_z$-$B$ critical line along $J_x=J_y$ at
$J_z=1$, for several $K$ between $3\times10^{-4}$ and $3\times10^{-2}$ (see
legend in panel~c). (a)~$|\delta|R_B$ versus $x=|\delta|/K^2$ in phase $B$.
(b)~$\delta R_{A_z}$ versus $x$ in phase $A_z$. (c)~$x\,[R_{A_z}(K,\delta)/R_{A_z}(0,\delta)-1]$
versus $x$. Thin horizontal lines in (b) and (c) mark the asymptotic values
$-4$ and $-128$ discussed in the text.}
\label{fig:scaling-collapse}
\end{figure*}

For the numerical test, we evaluate the Ricci scalar from Eq.~(\ref{gNK}) and
its first and second derivatives with respect to $(J_x,J_y)$, obtained by
differentiating the integrand analytically. The integrals are computed with
Gauss-Legendre quadrature on panels graded geometrically toward the merging
point. The results are converged to better than $10^{-3}$ under refinement,
reproduce the closed forms at $K=0$, and agree with the values quoted in
Eq.~(\ref{cross}). Figure~\ref{fig:scaling-collapse}(a,b) shows $|\delta|R$ as
a function of $x$ for $K$ between $3\times10^{-4}$ and $3\times10^{-2}$. For
$x\lesssim10$ the curves for $K=3\times10^{-4}$ and $10^{-3}$ agree to within
$1\%$ in phase $A_z$ and $4\%$ in phase $B$, and the limit $K\to0$ is approached
monotonically. The collapse is visibly worse for $K=3\times10^{-2}$. The values
$K=0.1$ and $0.15$ of Sec.~\ref{sec:finiteB} lie outside the scaling regime,
which explains why their crossover regions in Eq.~(\ref{cross}) look different.

Turning to the form of the scaling functions, we first consider small $x$
($|\delta|\ll K^2$). There one finds $\mathcal F_+>0$ and $\mathcal F_-<0$, that
is, $R_{A_z}>0$ and $R_B<0$ as in Eq.~(\ref{cross}). Both functions tend slowly
to zero, with local logarithmic slopes of about $0.25$ and $0.2$ at $x=0.03$.
This corresponds to a divergence of $|R|$ at the critical line that is weaker
than $1/\delta$. The data do not allow us to decide whether it is a power law,
and we do not quote an exponent.

At intermediate $x$, $\mathcal F_+$ has a maximum of about $1.05$ at
$x\simeq15$, and $\mathcal F_-$ has a minimum of about $-0.24$ at $x\simeq4$.
For the three smallest values of $K$, $R_{A_z}$ changes sign at
$x_+^*\simeq58$-$62$ and $R_B$ at $x_-^*\simeq39$-$58$. Both are compatible
with a common limit $x^*\approx60$, although $x_-^*$ converges more slowly. For
small $K$ the region in which the sign of $R$ does not identify the phase is
therefore $|\delta|\lesssim60K^2$. Its width grows quadratically, and not
linearly, with the field.

At large $x$ the two phases behave differently. On the gapped side the field is
a perturbation, and $R_{A_z}$ approaches the zero-field value of
Eq.~(\ref{RAz_crit}), i.e.\ $\delta R_{A_z}\to-4$. The leading correction is
quadratic in $K$, as expected for a metric that depends on the field only
through $\Delta_q^2$, but it has a large coefficient,
\begin{equation}
\frac{R_{A_z}(K,\delta)}{R_{A_z}(0,\delta)}-1\simeq-c\,\frac{K^{2}}{\delta},
\qquad c\approx1.3\times10^{2},
\label{Az-K2}
\end{equation}
for small $\delta$, as shown in Fig.~\ref{fig:scaling-collapse}(c). The
expansion parameter is thus $c/x$, so the perturbative regime starts only at
$x\gtrsim10^{3}$. The sign change at $x\approx60$ occurs where this expansion
has already broken down.

In phase $B$ at large $x$, Eq.~(\ref{RB_crit}) would predict
$|\delta|R_B\simeq\bar R_B|\delta|/\ln(1/K)$. At fixed $x$ this grows as
$K^{-1}x^{-1/2}$, so it cannot be part of a $K$-independent function. In the
whole range where the collapse holds, $R_B$ stays well below
$\bar R_B/\ln(1/K)$. The asymptotic form of Eq.~(\ref{RBK}) therefore describes
the vicinity of the critical line only at much smaller $K$ than we can resolve,
because its corrections are large and depend strongly on $\delta$. This is in
contrast to the point $(J_x,J_y)=(0.75,0.75)$ of Fig.~\ref{fig:fieldcut}(b),
which is far from the critical line. We therefore do not claim a scaling
collapse of $R_B$ in terms of $\bar R_B/\ln(1/K)$.

We emphasize the scope of these results. All statements in this subsection
refer to the diagonal $J_x=J_y$ and to the slice of fixed $K$ of
Sec.~\ref{sec:finiteB}. The values of $x^*$, $c$ and the extrema of
$\mathcal F_\pm$ are numerical results for $K\to0$. A derivation of
$\mathcal F_\pm$, including the exponents at small $x$, from the low-energy
theory near the merging point is left for future work.

\section{Discussion}
\label{sec:discussion}

The results of Secs.~\ref{sec:gapped}-\ref{sec:field} can be summarized as
follows. Away from the critical lines the curvature of the intensive metric is
negative in the Abelian phases and positive in the gapless phase (and, at finite
field, in the non-Abelian phase), and $|R|$ becomes large on the critical lines.
We now discuss what this curvature measures, how it can be accessed, and which
of these statements depend on conventions.

We start with the geometric meaning of the sign. At zero field the ground state
is a product of single-mode states, whose only parameter dependence is the phase
$\varphi_q(\lambda)$. The quantum metric, Eq.~(\ref{gN}), is therefore the
metric induced on the $(J_x,J_y)$ plane by the map
$\lambda\mapsto\{\varphi_q(\lambda)\}$ into a flat torus with line element
$\tfrac14\sum_q d\varphi_q^{2}$. The intrinsic curvature of a surface embedded
in flat space is fixed by its second fundamental form through the Gauss
equation, and nothing forces it to have a definite sign when the codimension
exceeds one. The sign is thus a property of how the family of ground states
bends inside the space of mode phases. That it is the same throughout each phase
and differs between them is a nontrivial result.

In the gapless phase the logarithmically divergent part of the metric is
determined entirely by the Dirac points. Let $\mathcal A$ be the area of the
triangle with sides $1$, $J_x$ and $J_y$ (Appendix~\ref{app:dirac}), for which
$-D=16\mathcal A^{2}$. The amplitude of Eq.~(\ref{RBbar}) can then be written as
\begin{equation}
\bar R_B=\frac{2\pi\,J_x^{2}J_y^{2}}{\mathcal A^{3}},
\label{RBarea}
\end{equation}
so the universal part of the curvature of phase $B$ depends on the couplings
only through the shape of this triangle. It diverges when the triangle becomes
degenerate, i.e.\ when the two Dirac points merge. The same function $D$
controls the gapped phases [Eqs.~(\ref{RAz}) and (\ref{D_Az})], where it is
positive and no triangle exists, and the curvature diverges when it vanishes
from that side.

It is also worth asking what the Ricci scalar adds to the metric itself. The
metric components are chart dependent: in the $A_x$ phase they differ from
those of the $A_z$ phase, with $g_{xx}\neq g_{yy}$ (Sec.~\ref{sec:gapped}). The
Ricci scalar removes this ambiguity and shows that the three gapped phases have
the same curvature function up to a change of coordinates, which cannot be seen
from the metric components alone. Moreover, the metric carries no sign, whereas
$R$ does, and its sign is what distinguishes the gapped from the gapless side in
the phase maps.

Next we come back to the exponents near the critical line, which are different
on the two sides of the transition. For the metric, a power-counting argument
near the $A_z$-$B$ line reproduces the observed behavior. The Dirac points
merge into a point with linear dispersion in one direction and quadratic in the
other, so the relevant momenta scale as $k_x\sim|\delta|^{1/2}$ and
$k_y\sim|\delta|$. Since $\partial_\delta\varphi\sim|\delta|^{-1}$, this gives
\begin{equation}
\frac{g}{N}\sim\frac{|\delta|^{3/2}}{|\delta|^{2}}=|\delta|^{-1/2},
\end{equation}
in agreement with $1/\sqrt{D}$.

The curvature exponents, $1$ for $R$ on the gapped side and $3/2$ for the
amplitude $\bar R_B$ on the gapless side [Eqs.~(\ref{RAz_crit}) and
(\ref{RB_crit})], do not follow from this counting alone. The reason is that the
metric becomes highly anisotropic at the critical line: $\cos(q^*_1-q^*_2)\to1$
and $g_{xx}\simeq g_{xy}\simeq g_{yy}$ at leading order. Its determinant then
stays finite while one eigenvalue diverges as $|\delta|^{-1/2}$ and the other
vanishes as $|\delta|^{1/2}$. A derivation of these exponents from the scaling
theory of Dirac-point merging is an open question. We also stress that
$\bar R_B$ is the coefficient of $1/\ln L$, so its divergence is defined with
the thermodynamic limit taken first and $\delta\to0$ afterwards. At fixed $L$
the divergence is cut off in the crossover region
$|\delta|\lesssim C(2\pi/L)^2$ of Sec.~\ref{sec:gapless}.

Regarding experimental access, with the normalization of Eq.~(\ref{gN}) the
metric is a spectral sum,
\begin{equation}
g_{\mu\nu}=\mathrm{Re}\sum_{n\neq0}
\frac{\langle0|\partial_\mu H|n\rangle\langle n|\partial_\nu H|0\rangle}
{(E_n-E_0)^2}
=\int_0^\infty\! d\omega\,\frac{S_{\mu\nu}(\omega)}{\omega^{2}},
\end{equation}
where $S_{\mu\nu}(\omega)$ is the spectral function of the operators
$\partial_\mu H$ \cite{kolodrubetz2013classifying}. In the present model
$\partial_{J_\alpha}H=-\sum_{\langle ij\rangle_\alpha}\sigma^\alpha_i\sigma^\alpha_j$
is a bond-energy operator. In the Majorana representation it is bilinear in the
itinerant fermions with fixed link variables, so it creates fermion pairs and no
fluxes, and the sum above stays in the vortex-free sector. Its spectral function
is of the type probed by Raman scattering on Kitaev materials. The weighting
$1/\omega^2$ emphasizes low energies, and the logarithm in phase $B$ reflects
the linear density of states of the Dirac fermions. Since $R$ is built from the
first and second derivatives of $g_{\mu\nu}$, it is in principle accessible from
the dependence of such low-energy spectral weight on the couplings, although we
do not expect this to be practical in current experiments.

The curvature of phase $B$ is defined only after the infrared divergence of the
metric is cut off. We used two cutoffs of different origin: the finite size $L$
in Sec.~\ref{sec:gapless}, and the field-induced mass in Sec.~\ref{sec:fieldIR}.
The latter yields the same universal amplitude with $\ln L\to\ln(1/K)$
[Eq.~(\ref{RBK})]. The universal content is therefore $\bar R_B$, whereas the
value of $R_B$ at a given $L$ or $K$ contains non-universal $O(1/\ln L)$
corrections that are numerically large [Fig.~\ref{fig:map}(b)]. The raw lattice
metric at finite $L$ is not suitable for this purpose, because the single mode
closest to a Dirac point produces resonances that are amplified in the curvature
(Sec.~\ref{sec:gapless}). A finite temperature would provide a third, physically
distinct cutoff, but this requires the Bures metric of a mixed state and is
beyond the present work.

Finally, several caveats should be kept in mind. All curvatures are computed
from the intensive metric $g_{\mu\nu}/N$. For the extensive metric the curvature
is smaller by a factor $N$ in every phase [Eq.~(\ref{Rscale})], so the signs,
the locations of the divergences and the exponents do not depend on this choice,
but the numerical values do. The ground state lies in the vortex-free sector
$W_p=+1$, so excitations that change $W_p$ and finite temperature are not
described. On a torus, the fermion parity of the state and the boundary
conditions determine whether the Majorana ground state projects onto a physical
spin state, which can affect the finite-size results in phase $B$.

At zero field the overall scale of the couplings does not change the ground
state, so $(J_x,J_y)$ at $J_z=1$ is a chart of a two-dimensional manifold and
the curvature is intrinsic. With the three-spin term, $K$ breaks this scale
invariance, and we study the slice of fixed $K$ in units of $J_z$. The scalar
curvature of a two-dimensional section of the three-parameter space generally
differs from that of other sections, so the results of Sec.~\ref{sec:field}
characterize this slice rather than the phases themselves. In addition,
Eq.~(\ref{HK}) gives $K\sim h_xh_yh_z/J^2$, so a fixed $K$ is not a fixed
magnetic field when $J_x,J_y$ are varied. The model with the three-spin term
should therefore be regarded as an exactly solvable extension of the Kitaev
model, and the larger values $K=0.1$ and $0.15$ lie outside the perturbative
regime of the field. Near the critical lines the sign of $R$ no longer
identifies the phase at finite field [Eq.~(\ref{cross})]; only the growth of
$|R|$ locates the transition there. These points define the range in which the
Ricci scalar is a reliable diagnostic: away from the crossover regions its sign
separates the gapped and gapless phases, and its divergence marks the critical
lines.

\section{Conclusion}\label{sec:conc}

In this work, we have used the Ricci scalar of the ground-state manifold to
study the phase transitions of the Kitaev honeycomb model. Starting from the
exact Majorana solution in the vortex-free sector, we reduced the ground state
to a product of two-level states in momentum space, derived the quantum
geometric tensor and the quantum metric on the two-dimensional manifold of
couplings $(J_x,J_y)$ with $J_z=1$, and computed the scalar curvature in all
four phases of the model.

In the gapped phases $A_x$, $A_y$, and $A_z$, we obtained closed-form
expressions for the metric and the Ricci scalar in the thermodynamic limit,
and showed that the three phases are related by an exact change of
coordinates on the ground-state manifold. In each of these phases the
curvature is negative. In the gapless phase $B$, the
Bloch function vanishes at two Dirac points, and the metric grows
logarithmically with the linear system size $L$. We showed that the raw
lattice metric at finite $L$ is dominated by resonances of single momentum
modes with the Dirac points, so that its curvature is not representative of
the macroscopic geometry, and we characterized phase $B$ through the smooth,
infrared-regularized metric. Its curvature is positive and scales as
$\bar{R}_B/\ln L$, with an $L$-independent amplitude $\bar{R}_B$ that
diverges as $|\delta|^{-3/2}$ at the phase boundaries. The change of sign of
the curvature between the gapped and gapless regions, together with its
divergence on the critical lines, allowed us to reconstruct the complete phase
diagram from the inverse scalar curvature, in full agreement with the known
phase structure of the model.

We have also included the three-spin interaction $K$ generated by a magnetic
field. The field gaps out phase $B$ and acts as an infrared cutoff, replacing
$\ln L$ by $\ln(1/K)$; the curvature of the non-Abelian phase is then finite
and positive in the thermodynamic limit and approaches $\bar{R}_B/\ln(1/K)$
as $K\to0$, while away from the critical lines the curvature of the Abelian
phases remains negative and depends on $K$ only through $K^2$. The critical
lines, which are not shifted by the field, are still marked by a strong growth
of $|R|$, but the field opens a crossover region around them, whose width
scales as $K^2$ for small $K$, in which the sign of $R$ no longer identifies
the phase.

These results show that the Ricci scalar is a sensitive and complementary probe
of quantum phase transitions. Since divergences of geometric quantities are
often regarded as signatures of phase transitions, the divergence of the Ricci
scalar provides a geometric signature of the critical lines in the Kitaev
honeycomb model. Moreover, its sign distinguishes the gapped and gapless phases
away from the crossover regions, while its critical scaling exhibits different
exponents on the two sides of the transition.

Several directions for future work follow naturally from this study. In the
presence of a field, a derivation of the scaling functions near the critical
line from the low-energy theory of Dirac-point merging, and of their relation
to the Berry curvature and the Chern number of the non-Abelian phase, would be
valuable. It would also be valuable to apply the same approach to other spin
models, such as the Kitaev chain and other exactly solvable or quantum-critical
spin models, in order to test whether the sign and divergence of the Ricci
scalar provide a general geometric signature of quantum phase transitions.

\appendix

\section{Dirac-point contribution to the metric}\label{app:dirac}

Near $q_D$ we write $q=q_D+\kappa$ and expand
\begin{equation}
h(q)\equiv1+J_xe^{iq_1}+J_ye^{iq_2}\simeq a_1\kappa_1+a_2\kappa_2\equiv w,
\end{equation}
where $a_\mu$ is given by
\begin{equation}
a_\mu=iJ_\mu e^{iq^*_\mu}.
\end{equation}
From Eq.~(\ref{dphi}),
\begin{equation}
\partial_\mu\varphi_q\simeq\mathrm{Im}\!\left(\frac{c_\mu}{w}\right),
\qquad
c_\mu=e^{iq^*_\mu}.
\end{equation}
The linear map $\kappa\mapsto w$ has Jacobian
\begin{equation}
\left|\mathrm{Im}(\bar a_1a_2)\right|=J_xJ_y\left|\sin(q_2^*-q_1^*)\right|,
\end{equation}
and the angular average of
$\mathrm{Im}(c_\mu/w)\,\mathrm{Im}(c_\nu/w)$ over $\arg w$ is
\begin{equation}
\frac{\mathrm{Re}(c_\mu\bar c_\nu)}{2|w|^2}.
\end{equation}
Integrating over an annulus between the infrared cutoff and a fixed outer
radius gives, per Dirac point,
\begin{equation}
\int\frac{d^2\kappa}{(2\pi)^2}\,\partial_\mu\varphi\,\partial_\nu\varphi
=\frac{\mathrm{Re}(c_\mu\bar c_\nu)}{4\pi J_xJ_y|\sin(q_2^*-q_1^*)|}
\ln\frac{1}{\epsilon}+O(1).
\end{equation}
The point $-q_D$ gives the same contribution.

Equation~(\ref{Dirac}) states that the complex numbers $1$, $J_xe^{iq_1^*}$,
and $J_ye^{iq_2^*}$ form a closed triangle with sides $1$, $J_x$, and $J_y$.
Its area is
\begin{equation}
\mathcal{A}=\tfrac12J_xJ_y\left|\sin(q_2^*-q_1^*)\right|,
\end{equation}
and Heron's formula gives $16\,\mathcal{A}^2=-D$, so that
\begin{equation}
J_xJ_y\left|\sin(q_2^*-q_1^*)\right|=\frac{\sqrt{-D}}{2}.
\end{equation}
Moreover, $|J_xe^{iq_1^*}+J_ye^{iq_2^*}|=1$ implies
\begin{equation}
\cos(q_1^*-q_2^*)=\frac{1-J_x^2-J_y^2}{2J_xJ_y}.
\end{equation}
Collecting the factor $1/8$ of Eq.~(\ref{gN}) and the two Dirac points yields
Eq.~(\ref{gB}).

\bibliography{refs}
\end{document}